\documentclass[12pt]{iopart}
\usepackage{epsfig}
\usepackage{iopams}
\newcommand{\ket}[1]{ {| #1 \rangle} }
\newcommand{\bra}[1]{ {\langle #1 |} }
\newcommand{\av}[1]{ {\langle #1 \rangle} }

\begin{document}

\title[Quantum master equation descriptions of a nanomechanical resonator]
{Quantum master equation descriptions of a nanomechanical
resonator coupled to a single-electron transistor}

\author{D. A. Rodrigues  and A. D. Armour}
\address{School of Physics and Astronomy, University of
Nottingham, Nottingham,\\ NG7 2RD, UK}
\ead{\mailto{denzil.rodrigues@nottingham.ac.uk},
\\\mailto{andrew.armour@nottingham.ac.uk}}

\begin{abstract}
We analyse the quantum dynamics of a nanomechanical resonator
coupled to a normal-state single-electron transistor (SET).
Starting from a microscopic description of the system, we derive a
master equation for the SET island charge and resonator which is
valid in the limit of weak electro-mechanical coupling. Using this
master equation we show that, apart from brief transients, the
resonator always behaves like a damped harmonic oscillator with a
shifted frequency and relaxes into a thermal-like steady state.
Although the behaviour remains qualitatively the same, we find
that the magnitude of the resonator damping rate and frequency
shift depend very sensitively on the relative magnitudes of the
resonator period and the electron tunnelling time. Maximum damping
occurs when the electrical and mechanical time-scales are the
same, but the frequency shift is greatest when the resonator moves
much more slowly than the island charge. We then derive reduced
master equations which describe just the resonator dynamics. By
making slightly different approximations, we obtain two different
reduced master equations for the resonator. Apart from minor
differences, the two reduced master equations give rise to a
consistent picture of the resonator dynamics which matches that
obtained from the master equation including the SET island charge.

\end{abstract}
\submitto{NJP}
\maketitle

\section{Introduction}

A very interesting class of nanoelectromechanical systems (NEMS)
is composed of a mesoscopic conductor, such as a quantum dot,
point contact, or single electron transistor, coupled
electrostatically to a nanomechanical
resonator~\cite{white,AB,mm,smirnov,set,mmh,ABZ,set2,oz,nems,Clerk1,ADA,clerk,ARzerof,milespre,rammer,SSET}.
Such devices have potentially important applications as
ultra-sensitive measuring devices~\cite{white,set,set2}, as well
as being interesting dynamical systems in their own right. A
question of particular interest for these systems is under what
circumstances the mechanical degrees of freedom require quantum
mechanics for their proper description.

When a resonator is coupled electrostatically to a mesoscopic
conductor as a mechanically compliant voltage-gate, the position
of the resonator affects the current flowing through the conductor
and hence the latter can be used to monitor the motion of the
resonator. However, the movement of electrons through the
conductor necessarily acts back on the resonator affecting its
dynamics in important ways. For weak electro-mechanical coupling,
the back-action of the conductor on the resonator is typically
analogous to an equilibrium thermal
bath~\cite{mm,mmh,ABZ,oz,Clerk1,clerk,milespre,rammer,SSET}. The
stochastic motion of electrons through the conductor gives rise to
a force on the resonator which leads to fluctuations in its state
that can be described by an effective temperature. Furthermore,
the conductor can also damp the oscillations of the resonator, so
that it reaches a thermal-like steady state\footnote{Recently, it
has been shown that more complex conductors like superconducting
SETs can also give rise to {\it negative} damping of the resonator
under certain circumstances~\cite{SSET}}.

The damping and heating effects of a conductor on a nearby
nanomechanical resonator were first reported by Mozyrsky and
Martin~\cite{mm} who examined the quantum dynamics of a resonator
coupled to a quantum point contact (QPC). In this
device~\cite{mm,Clerk1,rammer} the tunnelling matrix elements of a
tunnel junction depend linearly on the resonator position. The
back-action of the QPC on the resonator was obtained by tracing
over all the electronic degrees of freedom to derive a reduced
density matrix for the resonator. The resulting resonator master
equation was found to be of the Caldeira--Leggett form and hence
an effective temperature and damping constant arising from the
back-action of the electrons could be ascribed to the resonator.

The dynamics of a resonator coupled to a single electron
transistor (SET) has also been studied
theoretically~\cite{mmh,ABZ,oz,SSET}. In the simplest case,
electrons tunnel sequentially through a metallic island or quantum
dot gated by a mechanical resonator~\cite{ABZ,oz}. A classical
master equation description of this system was proposed for this
system in \cite{ABZ} and it was found that the resonator dynamics
could be described by a Fokker-Planck equation~\cite{milespre}
with the SET electrons again acting like a thermal bath. However,
apart from weak electro-mechanical coupling, it was also assumed
that the resonator moved very slowly on the time-scale of the
electron tunnelling time and that the energy associated with the
SET bias voltage was much greater than the resonator quanta. A
subsequent quantum mechanical study of the closely related system
of a resonator coupled to a quantum dot~\cite{oz} also showed that
the back-action of the electrons could give rise to resonator
damping, even without assuming a slow resonator or very high bias
voltage, but with a very different rate to that found in the
classical study.

In this paper, we begin from a microscopic description of the
SET-resonator system and proceed to derive reduced master
equations which describe the dynamics of the resonator alone. By
first tracing over the microscopic electronic levels, we obtain a
master equation which describes both the resonator and the excess
charge on the SET island. We show that although this master
equation is essentially equivalent to that proposed in the
classical description~\cite{ABZ}, it can nevertheless be used to
investigate the interesting question of what happens when the
resonator period matches the electronic tunnelling time. By
solving equations of motion for the resonator moments, we find
that the resonator is damped and reaches a thermal-like steady
state even when the resonator period is of order or less than the
electron tunnelling time, with maximum damping occuring when the
electrical and mechanical time-scales are equal.

Having obtained a master equation describing the SET island charge
and resonator, we then go on to derive, via different
approximations, two reduced master equations for the resonator
alone. Although the two reduced master equations have differences
in form, we find that (within their domains of validity) they both
lead to a description of the resonator dynamics which closely
matches that obtained from the master equation for the resonator
and SET island charge. In this framework, the two different
expressions for the resonator damping obtained before in
quantum~\cite{oz} and classical treatments~\cite{ABZ} emerge
naturally as limiting values for the cases where the average
tunnelling rate of the electrons is much less than the resonator
frequency, and the opposite case where the resonator frequency is
much smaller than the tunnel rate, respectively.

The outline of this paper is as follows. In section \ref{sec:meqn}
we outline the details of the nanomechanical-SET system we
consider and derive a master equation which describes the dynamics
of the resonator and the excess charge on the SET island for weak
electro-mechanical coupling. In section \ref{sec:numresults}, we
obtain equations of motion for the charge and resonator moments
and hence explore how the resonator dynamics depends on important
parameters such as the ratio of the relevant electrical and
mechanical time-scales. Then, in section \ref{sec:cl}, we trace
over the SET island charge to derive a reduced master equation for
the resonator alone, and compare the resulting resonator dynamics
with that obtained in the previous section.
 Next, in section \ref{sec:rwa}, we use a rotating
wave approximation to obtain an alternative reduced master
equation for the resonator which is valid in the limit where the
resonator frequency is much higher than the electron tunnelling
rate. Again we compare the resonator dynamics predicted by this
equation with the results of previous sections. Finally, in
section \ref{sec:conc} we present our conclusions and a brief
discussion of our results. The appendices contain more details on
certain aspects of the calculations.


\section{Master equation for resonator and SET island charge} \label{sec:meqn}

\begin{figure}
 \center{ \epsfig{figure=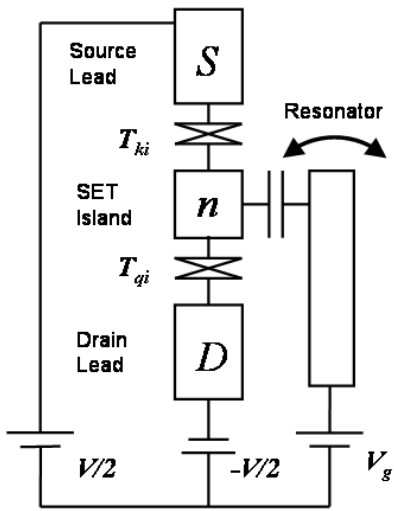,width=6cm}}
\caption{Schematic circuit diagram for the SET resonator system.}
\label{fig:schematic}
\end{figure}

The nanomechanical resonator-SET system we consider is shown
schematically in figure \ref{fig:schematic}. The SET consists of a
metallic island, with total capacitance $C_{\Sigma}$, connected
via tunnel junctions to two leads, with a bias voltage $V$ applied
across it. The nanomechanical resonator is adjacent to the SET
island and is coated with a metal layer so that it forms a
mechanically compliant voltage gate to which a voltage $V_g$ is
applied. Motion of the resonator modulates the gate capacitance
and hence the charging energy of the SET island, while changes in
the charge on the SET island modulate the equilibrium position of
the resonator. The resonator has a mass $m$ and is treated as a
single-mode harmonic oscillator with frequency $\omega_0$. Here we
are primarily interested in the effect of the SET on the resonator
and hence do not include any other influences on the resonator's
dynamics in our description.

Because the SET is metallic, the island as well as the source and
drain leads contain many (microscopic) electron energy levels. The
source (drain) lead has energy levels $\epsilon_{k(q)}$ and a
chemical potential $\mu_{S(D)}=-eV/2(+eV/2)$. The island has
energy levels $\epsilon_i$ and we set its chemical potential to
zero for simplicity. The resistances of the tunnel junctions,
$R_{S(D)}$, are taken to be much larger than the quantum
resistance, $R_Q=h/e^2$, so we can neglect higher order processes
such as co-tunnelling~\cite{Shnirman_98,schoeller}.

The dynamics of the resonator is affected by the overall charge on
the SET island, rather than the details of which of the
microscopic electron levels are occupied, hence we introduce the
macroscopic charge operator, $N$, for the total number of excess
electrons on the SET island. Using this macroscopic charge
variable, we can integrate over the microscopic electronic states,
whilst still keeping track of the overall island
charge~\cite{Shnirman_98}. The operator $\phi$, conjugate to $N$,
can be used to form the operators $e^{i \phi}$ and $e^{-i \phi}$
which increase and decrease the overall charge on the island
respectively.

The overall electrostatic energy of the SET island is determined
not just by the island charge, but also by the gate voltage (which
we write in dimensionless units as $n_g$) and the position of the
resonator. We assume that the electro-mechanical coupling is weak,
as has been the case in recent experiments~\cite{set,set2}, and
hence we consider only linear coupling between the SET island
charge and the resonator position~\cite{clerk}. Bearing these
details in mind, we write the Hamiltonian for this system
as~\cite{Shnirman_98,schoeller,oz}:
\begin{eqnarray}
H &=& \sum\limits_k \epsilon_k c_k^\dag c_k + \sum\limits_q
\epsilon_q c_q^\dag c_q + \sum\limits_i \epsilon_i c_i^\dag c_i
\nonumber\\&&+ {E_C}(N-n_g)^2 - \chi N(a^\dag +a)+ \hbar \omega_0 a^\dag a \nonumber \\
&&+ \sum\limits_k T_{ki}(c_k^\dag c_i e^{-i \phi} +c_k c_i^\dag
e^{+i \phi}) +\sum\limits_q T_{qi}(c_q^\dag c_i e^{-i \phi} +c_q
c_i^\dag e^{+i \phi}), \label{h1}
\end{eqnarray}
where $E_C=e^2/2C_{\Sigma}$ is the charging energy of the SET
island, $a$ is the resonator annihilation operator, $c_k,c_q$ and
$c_i$ are the annihilation operators acting on the electron levels
in the two leads and the island of the SET respectively, and
$T_{k(q)i}$ are the tunnelling matrix elements between the
microscopic states in the leads and the SET island. The coupling
between the SET island charge and the resonator is given by
$\chi=m \omega_0^2x_0x_q$, where $x_0$ is the shift in the
equilibrium position of the resonator when an electron is added to
the island (which in turn depends on details such as the gate
capacitance and voltage as well as the resonator-island separation
~\cite{ABZ}), and $x_q=\sqrt{\hbar/2m \omega_0}$ is the zero-point
position uncertainty of the resonator~\cite{nems}.

Assuming that the charging energy of the SET island, $E_C$, is
much greater than the thermal energy of the electrons, and that
the drain-source voltage is not too large (i.e.\ $eV/2\lesssim
E_{C}$), then only the two consecutive charge states $\ket{N_0},
\ket{N_1}$ closest to $n_g$ will be involved in the dynamics of
the system. In what follows we will restrict the charge state
basis to just these two states, and introduce the convenient
operators $n=\ket{N_1}\bra{N_1}$, $\sigma^+=\ket{N_1}\bra{N_0}$,
and $\sigma^-=\ket{N_0}\bra{N_1}$. Choosing the origin of the
resonator's position coordinate so that $\av{a^\dag +a}=0$
corresponds to the equilibrium position when there are $N_0$
electrons on the SET island, then the effective electrostatic
energy of the SET can then be written (discarding constants) as
$E_C(1-2n_g')$, where
$n_g'=n_g-N_0[1-\lambda^2\hbar\omega_0/E_C]$. The parameter $n_g'$
ranges from 0 to 1 and determines the difference in electrostatic
energy of the two charge states.

The form of the Hamiltonian (equation \ref{h1}) is, apart from the
leads, essentially that of the independent boson
model\cite{mahan,oz} and hence we make the canonical
transformation usually applied to such systems to eliminate the
term in the Hamiltonian describing the SET-resonator coupling. The
operators in this canonically transformed picture are given by
$\bar{A}=e^{-\lambda n (a^\dag - a)} A e^{\lambda n (a^\dag -
a)}$, where we have defined a dimensionless electromechanical
coupling parameter, $\lambda=\chi/\hbar \omega_0=x_0/(2x_q)$. The
canonically transformed Hamiltonian is then
\begin{eqnarray}
\bar{H} &=&\bar{H}_{S}+ H_B \nonumber\\
&&+\sum\limits_k T_{ki}\left[c_k^\dag c_i \sigma^-e^{\lambda
(a^\dag -
a)} +c_k c_i^\dag \sigma^+ e^{-\lambda (a^\dag - a)} \right]\nonumber\\
&&+\sum\limits_q T_{qi}\left [c_q^\dag c_i \sigma^- e^{\lambda
(a^\dag - a)} +c_q c_i^\dag \sigma^+ e^{-\lambda  (a^\dag - a)}
\right],
\end{eqnarray}
where the transformed system Hamiltonian is
\begin{equation}
\bar{H}_{S}=E_{ch} n + \hbar \omega_0 a^\dag a,
\end{equation}
with
\begin{equation} E_{ch}=E_C(1-2n_g')-\hbar \omega_0
\lambda^2,
\end{equation}
and the bath Hamiltonian describing the microscopic energy levels
on the metallic leads and the island is given by
\begin{eqnarray}
H_B&=&\sum\limits_k \epsilon_k c_k^\dag c_k + \sum\limits_q
\epsilon_q c_q^\dag c_q + \sum\limits_i \epsilon_i c_i^\dag c_i.
\end{eqnarray}
Converting to the interaction picture, $\bar{A}_I(t)={\rm
e}^{i(\bar{H}_S+H_B)t/\hbar}\bar{A}{\rm
e}^{-i(\bar{H}_S+H_B)t/\hbar}$, we obtain
\begin{equation} \label{eq:hint}
\bar{H}_I = \sum\limits_{l} T_{li}\left [c_l^\dag(t) c_i(t)
\sigma^-(t)e^{\lambda (a^\dag(t) - a(t))} +c_l(t) c_i^\dag(t)
\sigma^+(t) e^{-\lambda (a^\dag(t) - a(t))} \right],
\end{equation}
where the sum over $l$ runs over $k$ and $q$.

We wish to find an equation of motion for the density matrix of
our system, $\rho(t)$, which we define to be the resonator plus
the macroscopic charge degree of freedom. For high resistance
junctions, the tunnel couplings between the leads and the SET
island ($T_{li}$) are weak, thus we assume that the microscopic
energy levels in the leads and island remain in equilibrium
throughout with occupancies set by the Fermi-Dirac distribution
function (with the relevant chemical potential) so that
$\Xi(t)=\Xi(0)$ for all time, $t$, where $\Xi(t)$ is the density
matrix of the microscopic levels in the leads and the island (Born
approximation~\cite{oz,Gardiner,carmichael}). We also make the
conventional assumption that the full density matrix factorizes at
$t=0$, so that it takes the form $\rho(0)\Xi(0)$. Assuming that
the electrons in the leads relax on a time-scale much faster than
that over which $\dot{\bar{\rho}}_I$ evolves, we also make a
Markov approximation~\cite{carmichael} and hence arrive at a
master equation of the form,
\begin{eqnarray} \label{eq:bmapprox}
\dot{\bar{\rho}}_{I}(t) &=& -\frac{1}{\hbar^2} \int\limits_0^t dt'
{\rm Tr}\left[ \bar{H}_I(t), \left[\bar{H}_I(t'),
\bar{\rho}_I(t)\Xi \right]\right],
\end{eqnarray}
where the trace is over the microscopic electronic levels in the
leads and island.

Defining the bath operators $B_{il} = c_i c_l^\dag$ and the system
operator $S=\sigma^-e^{\lambda(a^\dag + a)}$, we can write the
equation of motion for the system density matrix as:
\begin{eqnarray}
\dot{\bar{\rho}}_{I}(t) &=& -\frac{1}{\hbar^2} \int\limits_0^t dt'
{\rm Tr}\sum\limits_{l,i}|T_{li}|^2\left[ B_{il}(t) S(t),
\left[B_{il}^\dag(t') S^\dag(t'),
\bar{\rho}_I(t)\Xi\right]\right]\nonumber \\
&&  -\frac{1}{\hbar^2} \int\limits_0^t dt' {\rm
Tr}\sum\limits_{l,i}|T_{li}|^2\left[ B_{il}^\dag(t) S^\dag(t),
\left[ B_{il}(t') S(t'), \bar{\rho}_I(t)\Xi \right]\right].
\end{eqnarray}
Notice that we need only sum over the microscopic levels once, as
only matched pairs of creation and annihilation operators
contribute when the electronic levels are in equilibrium.
 The time dependence of the system operator
$S(t)=\sigma^-(t) e^{\lambda(a^\dag(t) - a(t))}$ is somewhat
complicated, and and so we write this as a sum of eigenoperators
(defined by $[\bar{H}_{S}, S_m]=-\hbar\omega_mS_m$), each with a
time dependence of the form $e^{-i \omega_m t}$, where $\omega_m$
is an integer multiple of $\omega_0$. Anticipating the
approximation we will make later that the electro-mechanical
coupling is weak, we write out $S(t)$ as a series in $\lambda$:
\begin{eqnarray}
 S(t) &=& \sum_m \sigma^-(t)S_m e^{-i \omega_m t} \nonumber\\
&=& \sigma^-e^{-iE_{ch}t}\left[1 + \lambda a^\dag e^{i \omega_0
t}-\lambda ae^{-i \omega_0 t} +\frac{\lambda^2}{2}a^\dag a^\dag
e^{2i \omega_0 t}\right.\nonumber\\
&&\left.+\frac{\lambda^2}{2}a ae^{-2i \omega_0 t}
-\frac{\lambda^2}{2}a^\dag a-\frac{\lambda^2}{2}a a^\dag
+\cdots\right].
\end{eqnarray}
The operators $S_m$ are products of different numbers of the
operators $a$ and $a^\dag$, with $\omega_m$ determined by the
number of each. Putting the time dependence explicitly into the
equation of motion we obtain,
\begin{eqnarray} \label{eq:predelta}
\fl \dot{\bar{\rho}}_{I}(t) &=& -\frac{1}{\hbar^2} \int\limits_0^t
dt' {\rm Tr}\sum\limits_{l,i}|T_{li}|^2\sum\limits_{m,n} {\rm
e}^{-i (t-t')\left[\epsilon_i -\epsilon_l
+E_{ch}\right]/\hbar}{\rm e}^{-i (\omega_m t-\omega_nt')} \left[
B_{il} \sigma^-S_m, \left[B_{il}^\dag \sigma^+ S^\dag_n,
\bar{\rho}_I(t)\Xi \right]\right]\nonumber \\
\fl &&  -\frac{1}{\hbar^2} \int\limits_0^t dt' {\rm
Tr}\sum\limits_{l,i}|T_{li}|^2\sum\limits_{m,n}{\rm e}^{i
(t-t')\left[\epsilon_i -\epsilon_l +E_{ch}\right]/\hbar}{\rm
e}^{i( \omega_m t -\omega_nt')}\left[ B_{il}^\dag \sigma^+
S^\dag_m, \left[ B_{il} \sigma^-S_n, \bar{\rho}_I(t)\Xi \right]
\right].
\end{eqnarray}
We now assume that we can replace the integral over $t'$ with an
integral over $\tau=(t-t')$ and that the resulting integrand is
sufficiently peaked about $\tau=0$ that we can extend the upper
limit to infinity (assumptions consistent with the Markov
approximation~\cite{oz,Gardiner}). The time integrals can be
evaluated with the help of the expression: $\int\limits_0^\infty
d\tau {\rm e}^{i \omega \tau} =\pi
\delta(\omega)+iPV(1/\omega_0)$. The principal value terms lead to
coherent corrections to the evolution of the density matrix, but
since we are working in the limit where the junction resistances
are relatively high ($R_{S(D)}\gg R_Q$), these corrections are
small and can be neglected~\cite{oz,Shnirman_98,makhlin_01}.

Cyclicly permuting the bath operators under the trace, the master
equation can be rewritten in the form:
\begin{eqnarray}
\dot{\bar{\rho}}_{I}(t) &=&
 \frac{2\pi}{\hbar} {\rm
Tr}\sum\limits_{l,i}|T_{li}|^2\sum\limits_{m,n} \delta(\epsilon_i
-\epsilon_l +E_{ch}+\hbar\omega_n ) \left( B_{il}^\dag B_{il}
\mathcal{F}\left[\sigma^-S_n(t),  \sigma^+ S^\dag_m(t)\right]
\right.
\nonumber \\
&&\left. + B_{il}B_{il}^\dag \mathcal{F}\left[
 \sigma^+ S^\dag_n(t),
\sigma^- S_m(t) \right]\right)\bar{\rho}_I(t)\Xi,
\end{eqnarray}
where we have defined the super-operator,
\begin{eqnarray}
\mathcal{F}[X,Y]\rho = \frac{1}{2} (X\rho Y + Y^\dag \rho X^\dag)
-\frac{1}{2} ( Y X \rho + \rho X^\dag Y^\dag).
\end{eqnarray}

We now perform the trace over the bath, using ${\rm Tr} [c_k^\dag
c_k\Xi] = f(\epsilon_k -\mu_S)$, etc., and convert the sums over
the electron levels (i.e. over $k,q,i$) to integrals. After
integrating over $\epsilon_k$ and $\epsilon_q$, the master
equation becomes
\begin{eqnarray} \label{eq:fermi}
\fl \dot{\bar{\rho}}_I(t) &=& g_D\sum\limits_{m,n} \int d
\epsilon_{i}
f(\epsilon_i)[1-f(\epsilon_i+E_{ch}+\hbar \omega_n-\mu_D)] \mathcal{F}\left[\sigma^-S_n(t),  \sigma^+ S^\dag_m(t)\right] \bar{\rho}_I(t)\nonumber \\
\fl &+&g_D \sum\limits_{m,n} \int d \epsilon_{i}
[1-f(\epsilon_i)]f(\epsilon_i+E_{ch}+\hbar \omega_n-\mu_D) \mathcal{F}\left[\sigma^+ S_n^\dag(t),  \sigma^- S_m(t)\right] \bar{\rho}_I(t)\nonumber \\
\fl &+&g_S \sum\limits_{m,n} \int d \epsilon_{i}
f(\epsilon_i)[1-f(\epsilon_i+E_{ch}+\hbar \omega_n-\mu_S)] \mathcal{F}\left[\sigma^-S_n(t),  \sigma^+ S^\dag_m(t)\right] \bar{\rho}_I(t)\nonumber \\
\fl &+&g_S \sum\limits_{m,n} \int d \epsilon_{i}
[1-f(\epsilon_i)]f(\epsilon_i+E_{ch}+\hbar \omega_n-\mu_S)
\mathcal{F}\left[\sigma^+ S_n^\dag(t),  \sigma^- S_m(t)\right]
\bar{\rho}_I(t),
\end{eqnarray}
where the quantities $g_{S(D)}$ are defined by the relation
 \begin{equation}
 g_{S(D)}=\frac{1}{R_{S(D)}e^2} = \frac{2\pi}{\hbar} |T_{k(q)i}|^2
D_{k(q)}D_i,
\end{equation}
and it has been assumed that the tunnelling matrix elements
$T_{k(q)i}$ and the densities of states of the leads, $D_{k(q)}$,
and island, $ D_i$ are independent of energy over the relevant
range of $k(q),i$.

So long as the electron temperature in the leads is much less than
the other relevant energy scales, the Fermi-functions can be
replaced by simple step functions. Furthermore, for sufficiently
large bias voltages (such that $E_D=eV/2- E_{ch}
>\hbar \omega_n$ and $E_S=eV/2+ E_{ch}
>-\hbar \omega_n$) and sufficiently weak electro-mechanical coupling (we require $\lambda \ll 1$, so
that the series can be truncated after a finite number of terms),
tunnelling in the opposite direction to that set by the bias
voltage cannot occur and hence the master equation can be written
in the simplified form
\begin{eqnarray} \label{eq:meqn}
\dot{\bar{\rho}}_I(t) &=& g_S \sum\limits_{m,n}
(E_S+\hbar\omega_n) \mathcal{F}\left[\sigma^-S_n(t),  \sigma^+ S^\dag_m(t)\right]\bar{\rho}_I(t)\nonumber \\
&&+g_D \sum\limits_{m,n} (E_D-\hbar\omega_n)
\mathcal{F}\left[\sigma^+ S_n^\dag(t),  \sigma^- S_m(t)\right]
\bar{\rho}_I(t).
\end{eqnarray}
Converting the master equation  out of the interaction and
canonically transformed pictures, using $\sigma^-_{CT}=\sigma^-
e^{\lambda(a^\dag-a)}$ and $a_{CT}= a+\lambda n$, and working to
order $\lambda^2$, we find
\begin{eqnarray} \label{eq:cmeqn}
\dot{\rho}(t) &=& -\frac{i}{\hbar}\left[H_S, \rho(t) \right] \nonumber\\
 &&+g_S  \mathcal{F}\left[\sigma^-(E_S -\lambda\hbar\omega_0(a^\dag
 +a)  +\lambda^2\hbar\omega_0), \sigma^+ \right] \rho(t)\nonumber \\
&&+g_D \mathcal{F}\left[\sigma^+ (E_D+\lambda\hbar\omega_0(a^\dag
 +a)-\lambda^2\hbar\omega_0),  \sigma^-\right] \rho(t), \label{eq:n15}
\end{eqnarray}
where
\begin{equation}
H_S=E_C(1-2n_g')n + \hbar\omega_0 [a^\dag a-\lambda n(a^\dag+a)].
\end{equation}

The assumption that the electro-mechanical coupling is weak
compared to the resonator quantum, $\lambda\ll 1$ (i.e.\ $\chi \ll
\hbar\omega_0$) is the principal approximation made about the
resonator in deriving this master equation.  The derivation we
have given here is convenient for later calculation, but it is
also possible to derive equation (\ref{eq:cmeqn}) under the less
restrictive weak-coupling approximation, $\chi \ll eV$, as we
discuss in \ref{app:cmeqn}. By casting the master equation in its
Wigner function form (see \ref{app:cmeqn}), we see that it is
essentially equivalent to the classical one proposed in
\cite{ABZ}, even for intermediate voltages ($eV \gtrsim
\hbar\omega_0$) and fast oscillator motion ($\Gamma \ll 1$).

\section{Resonator and charge dynamics} \label{sec:numresults}

The quantum master equation we have derived for the SET island
charge and resonator, equation (\ref{eq:n15}), is readily
integrated numerically to give a complete description of the
combined dynamics of the resonator and the macroscopic SET charge.
However, for the normal state SET-resonator system we consider
(unlike its superconducting counterpart~\cite{SSET}) it is also
possible to obtain a closed set of equations of motion for the
moments of the electrical and mechanical degrees of
freedom\footnote{We did check by numerical integration that for
Gaussian initial states of the resonator, the master equation
preserves positivity (even though it is not of the Lindblad form)
and relaxes to a Gaussian steady state.}. In this section we use a
combination of analytical and numerical methods to investigate the
dynamical and steady state properties of the resonator and SET
moments. In later sections, we make further approximations in
order to derive more compact analytical descriptions of the
resonator dynamics and comparison with the results obtained in
this section play an important role in judging their fidelity.

We can calculate the equations of motion for the moments by
multiplying the master equation by the appropriate operator and
tracing over the resonator and charge states. Assuming for
simplicity that $g_S=g_D=g$, a closed set of equations for the
first moments is obtained,
\begin{eqnarray}
\frac{d}{dt}{\av{ n}} &=&
g\left[eV/2-E_C(1-2n_g')+\lambda\hbar\omega_0\av{a^\dag
+a}\right]-g eV \av{ n} \label{eq:nmoment}\\
\frac{d}{dt}{\av{a^\dag+a}} &=& i\omega_0
\av{a^\dag - a} \label{eq:xmoment}\\
\frac{d}{dt}{\av{a^\dag-a}} &=& i\omega_0 \left[ \av{a^\dag + a} -
 2\lambda \av{n}\right]. \label{eq:pmoment}
\end{eqnarray}
These equations of motion are naturally equivalent to those for a
classical system (ref.~\cite{ABZ}). Here we investigate their
behaviour over a wide range of parameter values and find in each
case that the resonator motion is closely analogous to that of a
damped harmonic oscillator with a shifted frequency. Of particular
interest is the effect of varying the ratio of electrical and
mechanical time-scales, $\Gamma=geV/\omega_0$, as only the regime
$\Gamma \gg 1$ was described for classical treatments of the
system~\cite{ABZ,milespre}.

\begin{figure}
 \center{ \epsfig{figure=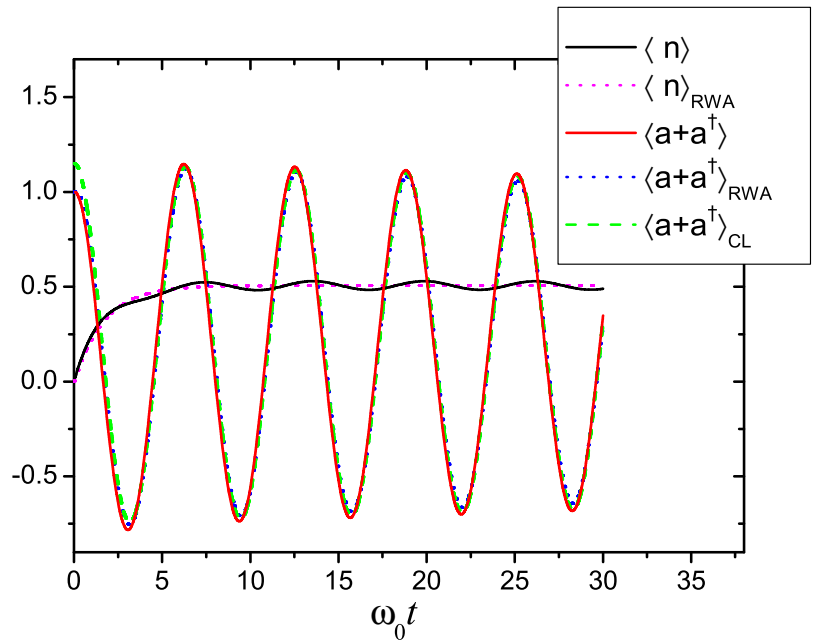,width=12cm}}
\caption{Evolution of the average resonator position,
$\av{a+a^\dag}$, and SET island charge, $\av{n}$. Results from a
numerical integration of the full master equation are compared
with results calculated using the Caldeira-Leggett (CL) and
Rotating Wave Approximation (RWA) master equations (described in
sections \ref{sec:cl} and \ref{sec:rwa}, respectively). Notice
that the Caldeira-Leggett curves used `slipped' initial
conditions~\cite{oppenheim} chosen to fit the long-time behaviour
of the numerical results. The parameters used are
$eV/\hbar\omega_0=6 $, $g=0.1/ \hbar$ (i.e.\ $\Gamma=0.6$) and
$\lambda=0.2$.} \label{fig:transient}
\end{figure}

Figure \ref{fig:transient} shows the time evolution of the moments
$\av{a+a^\dag}$ and $\av{n}$, obtained numerically, for a
particular choice of parameters. For the relatively fast
oscillator chosen ($\Gamma=0.6$) the average resonator position
shows a clear initial transient before decaying exponentially as
if it were a damped harmonic oscillator. The average SET island
charge, $\av{n}$, relaxes on two time-scales: initially, the
charge relaxes rapidly towards its steady state value over a time
$~geV$, it then undergoes weak oscillations (following those of
the resonator) which are damped out slowly.

\begin{figure}
 \center{ \epsfig{figure=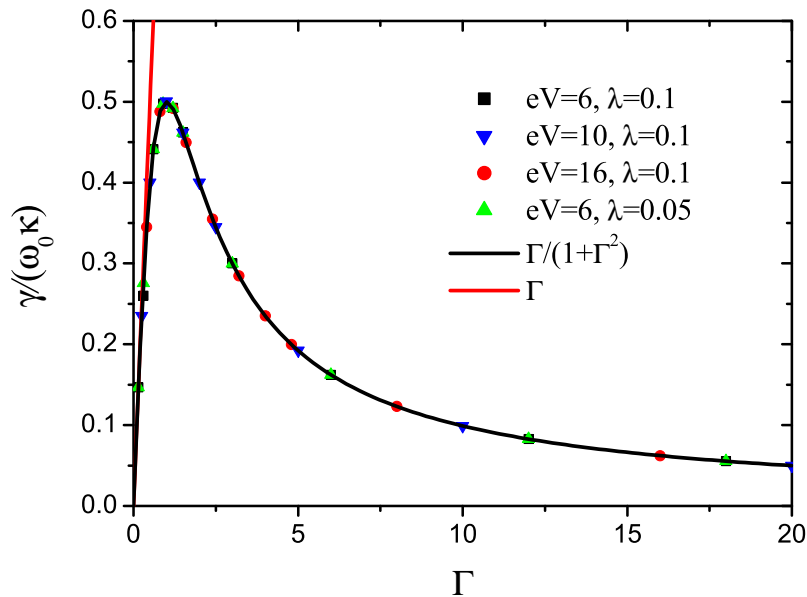,width=12cm}}
\caption{Plot of the resonator damping rate as a function of the
average tunnel rate, $\Gamma$.  For simplicity, the junction
resistances are assumed to be equal, $g_D=g_S=g$, and we choose
$n_g'=1/2$ (the degeneracy point). The voltages given for the
various data sets are measured in units of $\hbar \omega_0$. Also
shown are analytic curves which match expressions derived in
sections \ref{sec:cl} and \ref{sec:rwa}.} \label{fig:universal}
\end{figure}

\begin{figure}
 \center{ \epsfig{figure=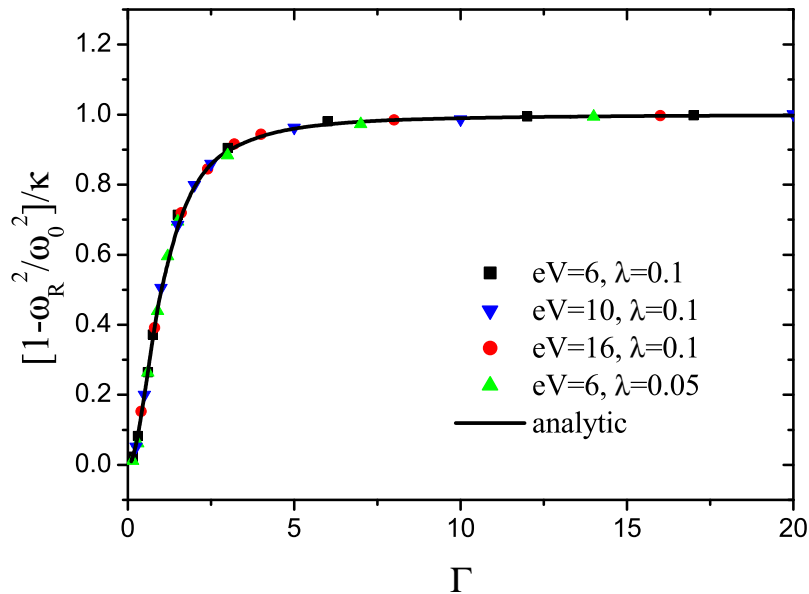,width=12cm}}
\caption{Plot of the relative frequency shift (squared) as a
function of $\Gamma$. The junctions are chosen to have equal
resistance $g_D=g_S=g$, and we choose $n_g'=1/2$ (the degeneracy
point). The voltages given for the various data sets are measured
in units of $\hbar \omega_0$. The analytic curve comes from
equation (\ref{eq:freqCL}) discussed in section \ref{sec:cl}.}
\label{fig:univ_freq}
\end{figure}

Although the dynamics of the resonator is qualitatively the same
for the whole range of parameters we have studied, there are
important quantitative differences. In figures \ref{fig:universal}
and \ref{fig:univ_freq} we plot the damping rate and frequency
shift (extracted from fits to the average resonator position as a
function of time) for different choices of $eV$, $\lambda$ and for
a wide range of $\Gamma$ values. Strikingly, we find that the data
all fall very close to universal curves. When time is measured in
units of the resonator period, the form of the universal curves
for damping and frequency shift are determined by $\Gamma$ and
$\kappa=2\lambda^2\hbar\omega_0/eV=m\omega_0^2x_0^2/eV$ which
gives a measure of the SET-resonator coupling strength compared to
to the bias voltage. As we shall see later on, the universal
curves emerge naturally from a reduced master equation for the
resonator.

From figure \ref{fig:universal}, we can see that the damping rate
reaches a maximum when the resonator period matches the tunnelling
rate, and drops to zero when the tunnelling rate is either much
faster or much slower than the oscillator. We also note that the
damping rate is proportional to $\Gamma$ for $\Gamma \ll 1$ and
inversely proportional for $\Gamma \gg 1$, with the latter
matching the results obtained in previous classical treatments of
the system~\cite{ABZ,milespre} while the former is consistent with
results found using a quantum treatment in \cite{oz}. The
frequency shift follows a different pattern, increasing
monotonically with $\Gamma$ until saturating at a constant value
for $\Gamma \gg1$.

Physically, the frequency shift and damping of the resonator arise
because the charge on the SET is a function of both the resonator
position and (indirectly) velocity and the resonator in turn
experiences a force proportional to the charge on the resonator.
Thus the resonator-SET coupling results in forces on the resonator
proportional to its position (frequency shift) and to its velocity
(damping) \cite{ARzerof}. The velocity dependent term is due to
the fact that the SET takes a finite amount of time to respond to
the resonator. If the SET responds instantly (which is equivalent
to taking the limit $\Gamma \to \infty$,) there is no damping.
Conversely, if the SET takes an essentially infinite amount of
time to respond to the resonator position (i.e.\ in the limit
$\Gamma \to 0$), then the damping must also disappear.

The second moments of the resonator can also be calculated from
equation \ref{eq:cmeqn}, and solved to find the steady state
variances of the resonator. We find,
\begin{eqnarray}
\frac{m\av{\delta u^2}_{ST}}{1-\kappa}=m\omega_0^2\av{\delta
x^2}_{ST}=eV\langle n \rangle_{ST}[1-\langle n \rangle_{ST}]
\label{eq:variances}
\end{eqnarray}
where
\begin{equation} \label{eq:nst}
\av{n}_{ST}=\frac{eV-2E_C(1-2n_g')}{2eV(1-\kappa)},
\end{equation}
is the steady state value of the island occupation (given by
equations \ref{eq:nmoment}-\ref{eq:pmoment}). In the weak-coupling
limit, $\kappa\to 0$, equipartition is recovered and if we also
take the high voltage limit, $eV\gg \hbar\omega_0$,
then\footnote{Away from the high voltage limit, the value of the
SET effective temperature will depend on the resonator frequency,
as pointed out in \cite{rammer}.} we can identify an effective
temperature for the resonator which matches previous classical
results~\cite{ABZ}: $k_BT_{SET}=eV\langle n \rangle_{ST}[1-\langle
n \rangle_{ST}]$. Note that these expressions for the steady state
variances are not valid for arbitrarily small voltages (and hence
do not violate the uncertainty relations) as they were derived
assuming a bias voltage large enough to ensure that no
back-tunnelling occurs. Specifically, equation
(\ref{eq:variances}) is valid for $eV>2(E_{ch}+2\hbar\omega_0$)
and $eV> 2(-E_{ch}+2\hbar \omega_0)$, and for weak
electro-mechanical coupling.

\section{Caldeira-Leggett Equation for the Resonator}
\label{sec:cl}

In section \ref{sec:meqn} we traced over the microscopic
electronic degrees of freedom to obtain a master equation for a
system consisting of the resonator and the macroscopic charge of
the SET island. A natural further step is to trace over the
remaining electronic degree of freedom and hence obtain a reduced
master equation for the resonator alone~\cite{milespre}. The
advantage of this reduced master equation is that it gives a
compact description of the resonator dynamics and makes clear the
analogy between the effect of the SET electrons and a thermal
bath.

In order to trace over the macroscopic charge variable, we treat
the SET as the bath and make many of the same kinds of assumptions
as were used in deriving equation (\ref{eq:cmeqn}) (details of the
calculation are given in \ref{app:cl}). In particular, we assume
that the tunnelling rate, $geV$, is fast compared to the rate of
change of the resonator's reduced density matrix in the
interaction picture (Markov approximation). Whilst we cannot be
completely sure when such an approximation is valid before
carrying out the calculation, we can judge its validity by
comparing the predictions of the resulting master equation with
those of equation (\ref{eq:cmeqn}) described in the previous
section. It is also necessary to assume that the coupling is weak
compared to the bias voltage (Born approximation), $\kappa \ll 1$,
but this is not restrictive given the weak-coupling assumption
used in the derivation of equation (\ref{eq:cmeqn}). We also
assume that the SET and resonator density matrices factorise.

The reduced master equation for the resonator we arrive at by
means of the Born-Markov approximations takes a very similar form
to that of the Caldeira-Leggett equation which describes quantum
Brownian motion~\cite{cl,paz1,tannor},
\begin{equation} \label{eq:cl}
\fl \dot{{\rho}}(t) = -\frac{i}{\hbar}[\tilde{H}_r, \rho_r(t)]
-\frac{i}{\hbar} \frac{\gamma_{CL}}{2} [x,\{p,\rho_r(t)\}]  -D
[x,[x,\rho_r(t)]]-\frac{f}{\hbar}[x,[p,\rho_r(t)]]
\end{equation}
where ${\tilde{H}}_r$ is the Hamiltonian of an harmonic oscillator
with the renormalised frequency,
\begin{equation}
\omega_R=\omega_0\left(1-\kappa
\frac{\Gamma^2}{1+\Gamma^2}\right)^{\frac{1}{2}} \label{eq:freqCL}
\end{equation}
and the damping constant is\footnote{For comparison, the damping
and renormalized frequency found in previous classical
calculations~\cite{ABZ,milespre,ARzerof} were
$\gamma=\kappa\omega_0/\Gamma$ and
$\omega_R=\omega_0(1-\kappa)^{1/2}$, respectively.}
\begin{equation}
\gamma_{CL}=\omega_0\frac{\kappa \Gamma}{1+\Gamma^2}.
\label{eq:gamaCL}
\end{equation}
The normal and anomalous diffusion constants~\cite{paz1,milespre}
are
\begin{equation}
D=\frac{m \gamma_{CL} eV}{\hbar^2}\langle n\rangle_{ST}[1-\langle
n\rangle_{ST}]
\end{equation}
and
\begin{equation}
\hbar f=-\frac{\kappa}{(1+\Gamma^2)}eV\langle
n\rangle_{ST}[1-\langle n\rangle_{ST}],
\end{equation}
respectively.

The steady state resonator variances are readily obtained from the
reduced master equation and hence we obtain
\begin{eqnarray}
m\omega_R^2\av{\delta x^2}_{ST} &=&m \av{\delta u^2}_{ST}-\hbar f\\
m\av{\delta u^2}_{ST} &=&eV\langle n\rangle_{ST}[1-\langle n\rangle_{ST}]\\
m\omega_0^2\av{\delta x^2}_{ST}&=& eV\langle n \rangle_{ST}
[1-\langle n\rangle_{ST}]
\frac{1+\frac{\kappa}{1+\Gamma^2}}{1-\frac{\kappa\Gamma^2}{1+\Gamma^2}}.
\end{eqnarray}
These values are somewhat different to those obtained before the
macroscopic charge variable was traced out. In particular, the
presence of the anomalous diffusion term means that the variances
in the position and velocity are no longer direcly proportional to
one another\footnote{The presence of an anomalous diffusion term
in the reduced master equation of a resonator coupled to a point
contact was reported in ~\cite{rammer}. However, in that case
there is no analogue of the resonator-SET charge master equations
which can be used to determine the `true' dynamics of the
resonator.}. Such differences must be interpreted as artifacts of
our Born-Markov approximations. However, in the limit $\kappa \to
0$ these values converge to the same limits as before.

Having compared the steady state properties of the resonator given
by the reduced master equation with the `true' values given by the
master equation for the resonator and SET island charge, we now
compare the dynamics. In figures \ref{fig:universal} and
\ref{fig:univ_freq} we plot the analytic Caldeira-Leggett values
of frequency shift and damping [equations (\ref{eq:freqCL}) and
(\ref{eq:gamaCL})] together with values extracted from numerical
integrations of the full master equation. The analytic results
match up very well with the numerics in the weak
electro-mechanical coupling limit. Furthermore, this good
agreement extends over the whole range of $\Gamma$ showing that
our Markov approximation remains valid even for $\Gamma\ll 1$.

 Although the reduced master equation successfully
describes the long-time dynamics of the resonator, it fails to
capture certain transient features which are present in the
numerics over time-scales of order the electron tunnelling time,
$geV$. In figure \ref{fig:transient} we compare the time evolution
of the moments calculated from equation (\ref{eq:cl}) to those
obtained from the full master equation. In order to match the long
time behaviour of the numerics we have had to `slip' the initial
conditions for the reduced master equation~\cite{oppenheim}. The
failure of a reduced master equation to capture transient features
in the motion of the system is often a consequence of Markov
approximations used in their derivation~\cite{oppenheim}, and it
seems very likely that the same has happened here.

An additional factor that may introduce further complications is
the presence of an external bath, which has been neglected in this
treatment. While naively we might expect that a thermal
environment could simply be incorporated into the reduced master
equation of the resonator (equation \ref{eq:cl}) by the addition
of further Caldeira-Leggett type terms~\cite{rammer}, the
correlation time of the environment provides an additional
timescale which might affect the validity of the assumptions used
in deriving the reduced master equation in the first
place~\cite{novotny}. A more detailed treatment would therefore
required, for instance including the environment at a more
microscopic level as a bath of harmonic oscillators in the
original Hamiltonian of the system.

\section{Rotating Wave Approximation} \label{sec:rwa}

An alternative way of deriving a reduced master equation for the
resonator alone is to make use of a rotating wave
approximation~\cite{Gardiner,sean} (RWA). In the RWA, terms with
an ${\rm e}^{i\omega_0 t}$ dependence are assumed to oscillate
very rapidly compared to the rate of change of $\bar{\rho}_I$ and
hence effectively average to zero. As with the validity of the
Markov approximation used in deriving the Caldeira-Leggett type
master equation, we cannot be sure in advance whether this
condition will be satisfied, but we can verify that it is {\it a
posteori} from the rate of change of $\bar\rho_I$ in the resulting
master equation.

Returning to equation (\ref{eq:meqn}), implementing the RWA means
dropping any term where $(\omega_m-\omega_n) \neq 0$. Expanding up
to order $\lambda^2$, the resulting master equation is
\begin{eqnarray} \label{eq:qmeqn}
\fl \dot{\bar{\rho}}_I &=& g_S E_S \mathcal{F}\left[\sigma^-,
\sigma^+\left\{1-\lambda^2(a^\dag
a+ a a^\dag)\right\}\right] \bar{\rho}_I \nonumber\\
\fl && +g_S \left\{ (E_S-\hbar\omega_0)\lambda^2\mathcal{F}\left[
a^\dag\sigma^-,  a \sigma^+\right]
+(E_S+\hbar\omega_0)\lambda^2\mathcal{F}\left[ a\sigma^-,
a^\dag\sigma^+ \right]
\right\}\bar{\rho}_I  \nonumber\\
\fl &&+g_D E_D\mathcal{F}\left[\sigma^+ ,
\sigma^-\left\{1-\lambda^2(a^\dag
a+ a a^\dag)\right\}\right] \bar{\rho}_I  \nonumber\\
\fl && +g_D \left\{ (E_D-\hbar\omega_0)\lambda^2\mathcal{F}\left[
a^\dag\sigma^+, a \sigma^- \right]
+(E_D+\hbar\omega_0)\lambda^2\mathcal{F}\left[a\sigma^+ ,
a^\dag\sigma^- \right] \right\}\bar{\rho}_I . \label{reduced}
\end{eqnarray}
This master equation is closely related to that found in~\cite{oz}
where a master equation for a closely related system of a quantum
dot gated by a nanomechanical resonator was derived.

When we make the simplifying assumption $g_S=g_D=g$, the dynamics
of the resonator and SET island charge decouple. The equation of
motion for the average charge then takes the simple form,
\begin{eqnarray}
\fl \frac{d}{dt}{\av{n}} = -g \left[E_S- \hbar
\omega_0\lambda^2\right]\av{n} + g\left[E_D- \hbar
\omega_0\lambda^2\right](1-\av{n}). \label{avn}
\end{eqnarray}
The behaviour of $\av{n}$ according to this equation is compared
with the `true' behaviour given by equations
(\ref{eq:nmoment})-(\ref{eq:pmoment}) in Figure
\ref{fig:transient}. It is clear that although the fixed point
value of $\av{n}$ remains the same, the dynamics is different as
it no longer displays oscillations correlated to the resonator
motion.

Assuming that the density matrix factorises into charge and
resonator parts, the evolution of the resonator alone is given by
a reduced master equation which has the Lindblad form
\begin{eqnarray} \label{eq:qmeq2}
 \dot{\bar{\rho}}_{r,I}&=& g \lambda^2 \left\{
\av{n}(E_S-E_D) + (E_D-\hbar\omega_0) \right\}
\mathcal{F}[a^\dag,a ] \bar{\rho}_{r,I} \nonumber \\
&&+g \lambda^2 \left\{ \av{n}(E_S-E_D) + (E_D+\hbar\omega_0)
\right\} \mathcal{F}[a,a^\dag ] \bar{\rho}_{r,I},
\end{eqnarray}
where
\begin{equation}
\bar{\rho}_{r,I}=\bra{N_0}\bar{\rho}_{I}\ket{N_0}+\bra{N_1}\bar{\rho}_{I}\ket{N_1}.
\end{equation}
The average charge appears in equation (\ref{eq:qmeq2}) as a
parameter though it is not necessary to assume that $\av{n}$ has
reached its fixed point value. Instead,  we can use the solution
of equation (\ref{avn}) to obtain a reduced master equation for
the resonator capable of capturing transient behaviour arising
from different initial conditions, albeit one with time dependent
coefficients.

The decoupling between the occupancy of the SET island and the
resonator motion occurs because we are working in the limit where
the resonator oscillations are much faster than the time-scale for
electron tunnelling events and hence the SET only experiences the
average of the resonator position. It should be noted, however,
that this decoupling only occurs in the strict RWA limit, i.e.\
$\Gamma \to 0$: for any non-zero values of $\Gamma$ there is a
residual correlation between the resonator motion and the average
charge. The oscillations in $\av{n}$ [obtained from equations
(\ref{eq:nmoment})-(\ref{eq:pmoment})] shown in Figure
\ref{fig:transient}, for which we set $\Gamma=0.6$,  clearly
follow those of the average resonator position, but this behaviour
disappears completely in the RWA limit.

Equations of motion for the moments of the resonator are readily
obtained from equation (\ref{eq:qmeq2}) or from equation
(\ref{eq:qmeqn}),
\begin{eqnarray}
\fl \frac{d}{dt}{\av{a^\dag + a}}  &=&-i\omega_0\av{a-a^\dagger}
-g \lambda^2 \hbar\omega_0
\av{a^\dag + a} +2g\lambda^3\hbar\omega_0\av{n}+2\lambda\frac{d}{dt} \av{n} \\
\fl i\frac{d}{dt}{\av{a^\dag -a}} &=&-\omega_0\av{a^\dagger+a}
+2\lambda \omega_0\av{n}-i g \lambda^2 \hbar\omega_0 \av{a^\dag-a
}.
\end{eqnarray}
In order to capture the long-time dynamics of the resonator as
simply as possible, it is sufficient to assume that the average
island charge has reached its steady state values. We can then
obtain the following equation of motion for the resonator position
about its fixed point value,
\begin{equation}
\av{\ddot{x}}=-\omega_0^2\av{x}-2g\lambda^2\hbar\omega_0\av{\dot{x}},
\end{equation}
where we have discarded terms ${\mathcal{O}}(\lambda^3)$ and
higher. From this equations it is clear that the resonator is
damped at a uniform rate
$\gamma_{RW}=2g\lambda^2\hbar\omega_0=\kappa\Gamma\omega_0$, but
there is no shift in its frequency. The steady state values of the
resonator variances take the form,
\begin{eqnarray}
m\omega_0^2\av{\delta
x^2}_{ST}&=&eV\av{n}_{ST}[1-\av{n}_{ST}]+\lambda^2m\omega_0^2x_q^2 \\
m\av{\delta
u^2}_{ST}&=&(eV-4\lambda^2m\omega_0^2x_q^2)\av{n}_{ST}[1-\av{n}_{ST}]+\lambda^2m\omega_0^2x_q^2.
\end{eqnarray}
As with the variances obtained from the Caldeira-Leggett master
equation, these differ slightly from the `true' values [equation
(\ref{eq:variances})], but they do approach the same weak-coupling
limit ($\lambda \to 0$).

In figure \ref{fig:transient} we compare the RWA evolution of the
system moments with the full case. We see that the agreement is
very close, including the short-time transient, for the value
$\Gamma=0.6$ used and in fact it becomes exact in the limit
$\Gamma \to 0$. The RWA values for the damping and temperature,
(and the absence of a frequency shift) are consistent with those
calculated from equations (\ref{eq:meqn}) and (\ref{eq:cmeqn}) in
the limit $\Gamma \ll 1$, as can be clearly seen from figures
\ref{fig:universal} and \ref{fig:univ_freq}. This implies that the
correct criteria for the validity of the rotating wave
approximation in this system is $\Gamma \ll 1$. This is in
contrast to the case of just a harmonic oscillator coupled to a
bath~\cite{Gardiner}, where the criteria for the validity of the
RWA is simply that the resonator frequency should be much larger
than the damping rate due to the bath. It is clear that the
presence of the SET in our system introduces an additional
time-scale over which $\bar{\rho}_I$ evolves and hence the period
of the resonator oscillations must be much shorter than both the
damping time and the electron tunnelling time for the RWA
approximation to be valid.

It is worth noting that equation \ref{eq:qmeqn} seems to suggest
that the damping is caused by terms like
$\mathcal{F}\left[a^\dag\sigma^+ , a \sigma^-\right]\bar{\rho}_I$
which add or remove an energy quantum from the resonator during a
tunnelling event (for discussions of energy exchange in similar
master equations see~\cite{oz,sean}). Although this mechanism
seems intrinsically quantum mechanical, involving the exchange of
energy in units of $\hbar \omega_0$, the RWA master equation was
derived from an equation describing essentially classical
behaviour. However, we must remember that the equation we have
derived is an {\it unconditional} master equation. Even though the
terms in equation~\ref{eq:qmeqn} describe the transfer of
individual quanta, the fact we are considering an ensemble of
systems means that the density matrix changes continuously. A full
description of damping that proceeds by transfer of discrete units
of $\hbar \omega_0$ (as equation~\ref{eq:qmeqn} seems to imply)
would require quantum trajectory methods~\cite{carmichael}.


\section{Conclusions and Discussion} \label{sec:conc}

Starting from a microscopic formulation, we have derived a quantum
master equation for the SET-resonator system which describes the
coupled dynamics of the SET island charge and the resonator in the
limit of weak electro-mechanical coupling and intermediate
voltages (i.e.\ $eV \gtrsim \hbar\omega_0$). By investigating the
dynamics of the charge and resonator moments we find that the
resonator is damped, undergoes a shift in frequency and eventually
reaches a thermal-like steady state. These results agree with
those found in a classical description of the same
problem~\cite{ABZ,milespre}, but are more general. In particular,
we find that the resonator is damped by the SET even when the
electron tunnelling time is of order the resonator period or
longer. Furthermore, maximum damping of the resonator by the SET
electrons occurs when the electron tunnelling time matches the
resonator period.

Starting with the resonator-island charge master equation, we have
derived reduced master equations for the resonator alone in two
different ways. The first reduced master equation is very similar
to the Caldeira-Leggett master equation and is obtained by tracing
over the island charge, treating it like an external bath (i.e.\
making Born-Markov approximations). The second type of reduced
master equation is found by making the rotating wave approximation
so that the charge dynamics decouples from the resonator motion.
Apart from an initial transient, the Caldeira-Leggett type master
equation describes the long-time dynamics of the resonator
faithfully over the whole range of relative electrical and
mechanical time scales. In contrast, the master equation obtained
via the rotating wave approximation only captures the resonator
dynamics when its period is much shorter than the electron
tunnelling time, though unlike the Caldeira-Leggett master
equation it captures the transient motion and is of the Lindblad
form so is guaranteed to preserve the positivity of the density
matrix.

 The derivation of a
reduced master equation for the resonator in the SET-resonator
system makes an interesting comparison with better known
dissipative systems as it proceeds via a two stage process:
tracing over the microscopic electronic levels gives rise to a
master equation which still contains the SET island charge, and
further approximations are needed to obtain a description of the
resonator alone. This is in contrast to other NEMS such as the
resonator-point contact system~\cite{mm,Clerk1,rammer} where
tracing over the microscopic electronic states leads directly to a
Caldeira-Leggett type master equation for the resonator. However,
the intermediate stage for the SET-resonator system (consisting of
a master equation for the resonator and SET island charge) leads
rather fortuitously to closed sets of equations of motion for the
resonator and charge moments. Thus it is possible to characterize
the dynamics of the resonator using simple numerical and
analytical techniques before making the additional approximations
required to obtain the reduced master equations of the resonator.
Thus we are able to show that the anomalous diffusion term which
arises in the Caldeira-Leggett type reduced master equation is not
a feature of the `true' resonator dynamics, but rather an artifact
of the approximations we make.

The quantum master equations we derive here for the SET-resonator
system will provide useful tools for further investigations into
the quantum dynamics of this system. In particular, we can use
them as a starting point for investigations of how an individual
SET and resonator (rather than an ensemble of such systems)
evolves. However, it will also be interesting to try to extend the
ensemble-averaged master equations derived here to obtain
descriptions of the low-bias and strong-coupling regimes where the
close connection between quantum and classical treatments found
here is unlikely to persist.

\ack We would like to thank S.D. Barrett, M.P. Blencowe and T.M.
Stace for useful comments and suggestions. We acknowledge
financial support from the EPSRC under grant GR/S42415/01.


\appendix

\section{Alternative derivation of the resonator and island
charge master equation}\label{app:cmeqn}

In the body of the paper we rewrote equation \ref{eq:meqn} to get
a master equation in a simple form (equation \ref{eq:cmeqn}).
Although the derivation of equation \ref{eq:meqn} required an
expansion in $\lambda$, this simple master equation is in fact
more general. Here we carry out a slightly different derivation
that does not require $\lambda\ll 1$, and is therefore valid in
the fully classical limit where both the coupling $\chi$ and the
bias voltage $V$ are much larger than $\hbar\omega_0$. Instead, we
make a different weak coupling approximation and assume that the
coupling is small compared to the bias voltage, $\chi\ll eV$. In
addition, we assume that the resonator moves slowly on the
relaxation time of the electrons in the leads.

We start from the Born-Markov master equation,
\begin{eqnarray*} \label{eq:bmapprox}
\dot{\bar{\rho}}_{I}(t) &=& -\frac{1}{\hbar^2} \int\limits_0^t dt'
{\rm Tr}\left[ \bar{H}_I(t), \left[\bar{H}_I(t'),
\bar{\rho}_I(t)\Xi \right]\right],
\end{eqnarray*}
where the interaction picture Hamiltonian (equation
\ref{eq:hint}), is
\begin{equation}
\fl \bar{H}_I = \sum\limits_{l} T_{li}\left[c_l^\dag(t) c_i(t)
\sigma^-(t)e^{\lambda (a^\dag e^{i\omega_0 t} - ae^{-i\omega_0 t}
)} +c_l(t) c_i^\dag(t) \sigma^+(t) e^{-\lambda (a^\dag
e^{i\omega_0 t} - ae^{-i\omega_0 t} )} \right].
\end{equation}
Using the Baker-Campbell-Hausdorff theorem~\cite{Gardiner} to
rearrange the terms under the integral so that the time dependence
in the exponentials is expressed in terms of $\tau=t-t'$, we then
assume that the resonator evolves slowly on the time-scale of the
bath correlation functions (i.e\ the relaxation time of the
electrons in the leads) and hence make the approximation
$\rm{e}^{i\omega_0\tau}\simeq 1+i\omega_0\tau$. Then we perform
the integration over $\tau$ to obtain
\begin{eqnarray}
\dot{\bar{\rho}}_{I}(t) &=& \frac{1}{\hbar^2} {\rm
Tr}\sum\limits_{l,i}|T_{il}|^2
B_{il}B_{il}^\dag\mathcal{F}\left[S'^\dag(t) e^{-\lambda(a^\dag -a)}, S'(t)\right.\nonumber \\
&& \left. \times\delta\left(\epsilon_i -\epsilon_l
+E_{ch}-\lambda\hbar\omega_0(a^\dag+a)-\lambda^2\hbar\omega_0\right)e^{\lambda(a^\dag
-a)} \right]\bar{\rho}_I(t)\Xi \nonumber\\
 &+&\frac{1}{\hbar^2} {\rm Tr}\sum\limits_{l,i}|T_{il}|^2
B_{il}^\dag B_{il} \mathcal{F}\left[S'(t) e^{\lambda(a^\dag -a)},S'^\dag(t)e^{-\lambda(a^\dag -a)}\right.\nonumber \\
&& \left.\times\delta\left(\epsilon_i -\epsilon_l
+E_{ch}-\lambda\hbar\omega_0(a^\dag+a)-\lambda^2\hbar\omega_0\right)
\right]\bar{\rho}_I(t)\Xi.
\end{eqnarray}
where
$S'(t)=\sigma^-e^{-i(E_{ch}/\hbar-\lambda\omega_0(a^\dag+a)-\lambda^2
\omega_0 )t}$. Although this appears unwieldy, converting back
from the interaction and canonically transformed picture using the
definition, $\rho(t)=
e^{\lambda(a^\dag-a)}e^{-iH_{S}t}\rho_I(t)e^{+iH_{S}t}e^{-\lambda(a^\dag-a)}$,
the above equation reduces to a simple form. After tracing over
the bath, performing the integrals over the Fermi functions
(following the same approach as in section \ref{sec:meqn}), and
assuming that the electro-mechanical coupling is sufficiently weak
to ensure that no back-tunnelling occurs, we regain equation
\ref{eq:cmeqn}:
\begin{eqnarray*}
\dot{\rho}(t) &=& -\frac{i}{\hbar}\left[H_S, \rho(t)
 \right] +g_S  \mathcal{F}\left[\sigma^-(E_S -\lambda\hbar\omega_0(a^\dag
 +a)  +\lambda^2\hbar\omega_0)
, \sigma^+\right] \rho(t)\nonumber \\
&&+g_D \mathcal{F}\left[\sigma^+ (E_D+\lambda\hbar\omega_0(a^\dag
 +a)-\lambda^2\hbar\omega_0),  \sigma^-\right] \rho(t).
\end{eqnarray*}
Note that in this case the condition on the coupling required to
ensure that no back-tunnelling occurs is $\lambda\ll
eV/\hbar\omega_0$ (i.e\ $\chi \ll eV$).

Writing the charge-diagonal matrix elements
$\bra{N_0}\dot{\rho}\ket{N_0}$ and $\bra{N_1}\dot{\rho}\ket{N_1}$
of equation (\ref{eq:n15}) in Wigner function form~\cite{Gardiner}
results in a pair of coupled equations equivalent to the classical
master equations proposed in ref.~\cite{ABZ}
\begin{eqnarray} \label{eq:wigner}
\fl \dot{W}_{00}(x,u)&=&  \left( \omega_0^2
x\frac{\partial}{\partial u}-u \frac{\partial}{\partial x} \right)
W_{00}(x,u)+g_S\left[E_S-m\omega^2x_0x+\lambda^2\hbar\omega_0\right]
W_{11}(x,u)\nonumber\\\fl  &&
-g_D\left[E_D+m\omega^2x_0x-\lambda^2\hbar\omega_0\right]
W_{00}(x,u) \\
\fl \dot{W}_{11}(x,u)&=& \left(\omega_0^2
(x-x_0)\frac{\partial}{\partial u}-u \frac{\partial}{\partial x}
\right)
W_{11}(x,u)-g_S\left[E_S-m\omega^2x_0x+\lambda^2\hbar\omega_0\right]
W_{11}(x,u)\nonumber\\
\fl && +g_D\left[E_D+m\omega^2x_0x-\lambda^2\hbar\omega_0\right]
W_{00}(x,u).
\end{eqnarray}

\section{Caldeira-Leggett equation for Resonator}\label{app:cl}
It has been shown~\cite{milespre} that when the resonator coupling
is small compared to the gate voltage ($\chi\ll eV$), and the
separation of timescales is large ($\Gamma \gg1$), a further
Born-Markov approximation can be performed on the classical master
equation~\cite{ABZ}, to get a Fokker-Planck equation for the
resonator alone. We can perform a closely analogous procedure,
starting with equation \ref{eq:n15}, to get a master equation for
the resonator alone, but we do not make the assumption $\Gamma \gg
1$.

First, we note that the diagonal elements of the SET
$\rho_{00}=\bra{N_0}\rho\ket{N_0}$ and
$\rho_{11}=\bra{N_1}\rho\ket{N_1}$ decouple from the off-diagonal
terms. Writing the density matrix as a vector
$\rho=\left(\rho_{00}, \rho_{11} \right)$, we can re-write
equation \ref{eq:n15} as,
\begin{eqnarray}
\dot{\rho} &=& -\frac{i}{\hbar} [ H_r, \rho ] - i\omega_0
\left(\frac{\av{a^\dag+a}_{ST}}{2} - \lambda
\left(\begin{array}{cc}  0 & 0 \\  0 & 1 \\\end{array}\right) \right)  [(a^\dag+a), \rho]\nonumber \\
&&+ g\left( \begin{array}{cc}  -E_D' & E_S' \\  E_D' & -E_S' \\
\end{array}\right) \rho +\frac{g\lambda\hbar\omega_0}{2} \left(\begin{array}{cc}
  -1 & -1 \\  1 & 1 \\\end{array}\right) \left\{(a^\dag+a),
  \rho\right\}, \label{eq:B1}
\end{eqnarray}
where $H_r=\hbar\omega_0 a^\dag a$, the curly braces, $\{.,.\}$,
indicate an anti-commutator and we have shifted the origin of the
oscillator position by the steady state value $\av{a^\dag+a}_{ST}$
so that the new fixed point will be zero. The electrostatic energy
differences associated with tunnelling are given by
\begin{eqnarray}
\fl E_{D}'&=&E_{D} -\left(\lambda^2\hbar\omega_0+ \lambda
\hbar\omega_0\av{a^\dag+a}_{ST}\right)=\frac{eV}{2}-E_{C}(1-n'_g)+
\lambda
\hbar\omega_0\av{a^\dag+a}_{ST}\\
\fl E_{S}'&=&E_{S} +\left(\lambda^2\hbar\omega_0+ \lambda
\hbar\omega_0\av{a^\dag+a}_{ST}\right)=\frac{eV}{2}+E_{C}(1-n'_g)-
\lambda \hbar\omega_0\av{a^\dag+a}_{ST}.
\end{eqnarray}

To obtain equation \ref{eq:n15}, we traced over the electron
levels to get an effective master equation for the resonator and
the charge on the SET island. Now, we wish to trace over the
remaining charge degree of freedom to leave a master equation for
just the resonator. We switch to an interaction picture
representation defined by $\tilde{\rho}(t)= e^{gH_{SET}t}e^{-iH_r
t/\hbar}\rho(t) e^{iH_r t/\hbar}$, where
\[
H_{SET}=\left( \begin{array}{clcl}  -E_D' & E_S' \\  E_D' & -E_S'
\end{array}\right),
\]
is the non-Hermitian `Hamiltonian' that describes the non-unitary
evolution of the SET in the absence of interaction with the
resonator. We define the SET-only operators as
\begin{eqnarray}
B_1&=&\frac{g \lambda\hbar\omega_0}{2} \left(\begin{array}{clcl}
  -1 & -1 \\  1 & 1 \end{array}\right)\\
B_2&=&\hbar\omega_0 \left(\frac{\av{a^\dag+a}_{ST}}{2} -
  \lambda
\left(\begin{array}{clcl}  0 & 0 \\  0 & 1\end{array}\right)
\right).
\end{eqnarray}
In this interaction picture, the equation of motion for the
density matrix is
\begin{eqnarray}
\dot{\rho}_I(t) &=& -\frac{i}{\hbar} B_2(t)[(a^\dag+a)(t),
\rho_I(t)] +B_1(t)\{(a^\dag+a)(t),\rho_I(t)\}.
\end{eqnarray}
Following a very similar procedure to that used to obtain equation
(\ref{eq:bmapprox}), we perform a Born-Markov
approximation~\cite{milespre} and hence obtain an equation of
motion for the resonator alone,
\begin{eqnarray}\label{eq:bmcl2}
\fl \dot{\tilde{\rho}_{rI}}(t) &=&
-\frac{1}{\hbar^2}\int\limits_0^t {\rm
Tr}{B}_2(t){B}_2(t')P_{SET}(0)[(a^\dag+a)(t),[(a^\dag+a)(t'),
{\rho}_I(t)]] \; dt'\nonumber\\
\fl && -\frac{i}{\hbar}\int\limits_0^t {\rm
Tr}{B}_2(t){B}_1(t')P_{SET}(0)[(a^\dag+a)(t),\{(a^\dag+a)(t'),
{\rho}_I(t)\}] \; dt'\nonumber\\
\fl && -\frac{i}{\hbar}\int\limits_0^t {\rm
Tr}{B}_1(t){B}_2(t')P_{SET}(0)\{(a^\dag+a)(t),[(a^\dag+a)(t'),
{\rho}_I(t)]\} \; dt'\nonumber\\
\fl && +\int\limits_0^t {\rm
Tr}{B}_1(t){B}_1(t')P_{SET}(0)\{(a^\dag+a)(t),\{(a^\dag+a)(t'),
{\rho}_I(t)\}\} \; dt'
\end{eqnarray}
where $P_{SET}$ is a vector that describes the occupation of the
SET. The Born approximation is based on the assumption of weak
electro-mechanical coupling which from equation (\ref{eq:B1})
implies $\hbar\omega_0\lambda^2\ll eV$ (i.e.\ $\kappa\ll 1$). The
Markov approximation assumes that the SET correlation functions
decay very rapidly on the time-scale of the rate of change of the
interaction picture density matrix~\cite{carmichael}.

We now need to evaluate the SET correlation functions. The only
non-zero terms are,
\begin{eqnarray}
{\rm Tr}[ B_2(t)B_1(t')
P_{SET}(0)]&=&-\frac{g(\hbar\omega_0)^2\lambda^2}{2}
e^{-geV(t-t')}\\
{\rm Tr} [B_2(t)B_2(t')
P_{SET}(0)]&=&{(\hbar\omega_0)^2\lambda^2}\frac{E_D'E_S'}{eV}
e^{-geV(t-t')},
\end{eqnarray}
and, in line with our Markov approximation, we discard terms that
decay like $e^{-geVt}$\footnote{It is therefore no surprise that
the equation we eventually obtain shows transient deviations from
the `true' resonator behaviour on times $\sim geV$.}. Noting that
$(a^\dag+a)(t)=(a^\dag+a)\cos(\omega_0t)+i(a^\dag-a)\sin(\omega_0t)/\omega_0$,
we can insert the correlation functions into equation
\ref{eq:bmcl2} and do the integrals over $t'$ exactly. Converting
back to the Schr\"{o}dinger picture,
\begin{eqnarray}
\dot{{\rho_r}}(t) &=& -\frac{i}{\hbar}[H_r, \rho_r(t)] \nonumber\\
&&-\frac{1}{\hbar^2}  \left(
\lambda^2\hbar^2\omega_0\frac{E_D'E_S'}{(eV)^2}\right)\frac{
 [(a^\dag+a),[\Gamma(a^\dag+a)-i(a^\dag-a),
{\rho}_r(t)]]}{\Gamma^2+1} \nonumber\\
&& +\frac{i}{\hbar}\frac{g \lambda^2\hbar^2\omega_0}{2}\frac{
[(a^\dag+a),\{\Gamma(a^\dag+a)-i(a^\dag-a),
{\rho}_r(t)\}]}{\Gamma^2+1},
\end{eqnarray}
we find a master equation that can be easily rearranged into a
more familiar form, which is very close to the Caldeira-Leggett
equation,
\begin{eqnarray} \label{eq:CLapp}
\dot{{\rho}}(t) &=& -\frac{i}{\hbar}[H_r,
\rho_r(t)]-\frac{i}{\hbar}\left(\frac{g\lambda^2\hbar^2\omega_0}{2x_q^2}\frac{\Gamma}{1+\Gamma^2}\right)[x^2,
\rho_r(t)]  -\frac{i}{\hbar} \frac{\gamma_{CL}}{2}
[x,\{p,\rho_r(t)\}]\nonumber
\\&& -D
[x,[x,\rho_r(t)]]-\frac{f}{\hbar}[x,[p,\rho_r(t)]].
\end{eqnarray}
The second and first terms can be combined, leading to a
renormalization of the resonator frequency,
\begin{equation*}
\omega_R=\omega_0\left(1-\kappa
\frac{\Gamma^2}{1+\Gamma^2}\right)^{\frac{1}{2}}
\end{equation*}
and the damping, diffusion and anomalous diffusion constant are
given by
\begin{eqnarray*}
\gamma_{CL}&=&\frac{2\lambda^2g\hbar\omega_0}{\Gamma^2+1}=\omega_0\frac{\kappa
\Gamma}{1+\Gamma^2}\\
D&=&\frac{m\gamma_{CL}E_D'E_S'}{eV\hbar^2}\\
 \hbar f&=&-\frac{E_D'E_S'\kappa}{eV(1+\Gamma^2)},
\end{eqnarray*}
 respectively. These
values are consistent with previous (classical) calculations in
the limit $\Gamma \gg1$~\cite{ABZ, milespre}.

\section*{References}


\end{document}